\documentstyle[prl,aps,epsf]{revtex}
\begin{document}
\twocolumn[\hsize\textwidth\columnwidth\hsize\csname@twocolumnfalse\endcsname
\title{
DECORATION OF JOSEPHSON VORTICES BY PANCAKE VORTICES IN
Bi$_{2}$Sr$_{2}$CaCu$_{2}$O$_{8+\delta}$. }
\author{V.\ K.\ Vlasko-Vlasov$^{1}$, A.\ E.\ Koshelev$^{1}$,  U.\ Welp$^{1}$,
G.\ W.\ Crabtree$^{1}$, and K.\ Kadowaki¨$^{2}$}
\address{
$^{1}$Materials Science Division, Argonne National Laboratory, 9700 South Cass Avenue, Argonne ,
IL 60439\linebreak[1]
$^{2}$Institute of Materials Science, The University of Tsukuba, 1-1-1 Tennodai, Tsukuba, 305-8573,
Japan}
\date{\today}
\maketitle
\begin{abstract}
Josephson vortices are imaged magneto-optically due to their
decoration with pancake vortices in
$Bi_{2}Sr_{2}CaCu_{2}O_{8+\delta}$ single crystals. Peculiarities
of interaction between the pancake and Josephson vortices (JV)
depending on the values of crossing fields and temperature are
studied based on the observations of these decoration patterns.
Evidences of the period-doubling in the decoration patterns
compared to the JV stack period, migration of JV lines between
neighboring stacks, and transitions between different JV
configurations are reported. Imaging of the pancake/Josephson
vortex decoration patterns over large areas is shown to be a
sensitive tool for detecting local variations of the anisotropy
and mapping imperfections in layered HTS samples.

\end{abstract}
\vskip1pc]
\narrowtext

\section{Introduction}
In strongly anisotropic high-T$_{c}$ superconductors, such as
$Bi_{2}Sr_{2}CaCu_{2}O_{8+\delta}$ (BSCCO), different magnetic
flux components are carried by different types of the magnetic
vortices: pancake vortices residing in the
 cuprate planes are responsible for the perpendicular flux component and Josephson vortices
 centered between the cuprate planes present the in-plane flux
 \cite {[1],[2],[3],[4],[5]}. In the first approximation,
 these two types of vortices are not interacting and form crossing lattices when the external
 field is not very close to the {\bf ab}-plane \cite {[6],[7]}. In this state pancake vortices form a conventional
 hexagonal lattice which coexists with a stretched hexagonal lattice of Josephson vortices.
Due to the large anisotropy $\gamma$ ($\sim$500 for BSCCO) the
latter arranges into stacks of Josephson lines
 with large distances between the stacks along the cuprate planes and $\gamma$-times smaller
  inter-vortex separation in the perpendicular direction. It was predicted that the magnetic
    flux should lock into the $ab$ plane at smaller field angles (to be more precise, if the normal
    field component is small enough, $H_{z}<H_{c1}^{c}$) and form a lattice of tilted vortices when the
    field is far from the plane \cite {[6]}. However, the experiment revealed a more complicated picture
    that can not be explained neglecting the interactions between pancakes and Josephson vortices.
     An account of these interactions results in a rich phase diagram of various vortex states in
     inclined fields \cite {[8],[9],[10],[11]} where crossing lattices present the ground state of the superconductor
     in a wide range of field angles from orientations close to the
     $ab$ plane to those close to the $c$ axis \cite {[9]}.

Already in the early decoration experiments in BSCCO under inclined fields the formation of vortex
rows along the in-plane field has been observed \cite {[12],[13]}. They were initially treated as chains of tilted
 vortices in accordance with the theory for the moderately anisotropic
 superconductors \cite {[14],[15],[16]}.
 In such materials, unlike in the isotropic case, currents around vortices tilted from the anisotropy
 directions redistribute so that a magnetic field changes sign on one side of the tilted vortex at some
 distance from its axis. This field, although weak, attracts neighboring vortices in the tilt plane thus
 forming chains of inclined vortices. It turned out, however, that in HTS with a weak Josephson
 coupling between the cuprate layers the scenario is more intriguing. Calculations showed that
 pancakes and Josephson vortices coexisting in tilted fields attract each other and as a result pancake
 stacks {\it decorate} Josephson vortex stacks \cite {[9]}. A similar explanation of the observed vortex chains
 in BSCCO, although without supporting calculations, was first suggested
 in \cite {[17]}. This decoration
 effect allows to image otherwise invisible (on the $ab$ plane) Josephson vortices and study their
  arrangement and changes in layered superconductors (at least in the range of temperatures
  where the interaction between the Josephson vortices and pancakes is larger than pinning by defects).
   Recent vortex imaging using scanning Hall probe microscopy gave direct indication for this scenario
   and for different phases of the interpenetrating pancake and Josephson
   vortex systems \cite {[18]}. Also, formation of pancake vortex chains and enhanced mobility of pancakes along
   the Josephson vortices have been observed recently by the transmission electron microsopy\cite{MatsudaSci02}

   In the present study we utilize high-resolution magneto-optical imaging \cite {[19]} to investigate
   the evolution of the coupled pancake (PV) / Josephson (JV) vortex system in dependence on
   temperature, applied field, and imperfections in single crystal BSCCO plates. It is revealed
   that PV/JV coupling is relatively strong and results in a joint motion of both vortex types
   under the direct action on either of them. We show that in addition to chains of pancake
   vortices on each JV line at small normal fields there can be superstructures with commensurate
   decoration patterns (each second JV stack occupied by PVs) as suggested
   in \cite {[20]}. Also,
   evidence for formation and motion of ``dislocations''and ``interconnects'' in the Josephson
   lattice in increasing field is presented.  A possibility of transitions between two JV lattice
   configurations resulting in steps on the field dependence of the JV stack period is discussed.
    We point out that the spacing of Josephson vortex stacks and their mobility is sensitive to the
    local density of pancakes and the defect structure in the sample. The latter results in spatial
    variations of the anisotropy and considerable bending of the JV lines.

\section{Experiment and Discussion}
The samples studied here were BSCCO single crystals with as grown
flat surfaces and straight edges. $T_{c}$ of the samples was
$\sim$ 90K as determined from the disappearance of the
 magneto-optical contrast at the sample edge.  The fields parallel and perpendicular to the
  sample plane were produced by two separate sets of coils positioned outside the optical cryostat,
  which allowed varying the field components independently. For imaging the normal component
   of the magnetic induction on the sample surface the magnetic garnet indicators were used as
   described in a recent review \cite {[19]}. The indicator film was placed on a flat sample surface and
   variations of the normal field, $B_{z}$, were revealed as local changes of the image intensity in
    the polarized light microscope. The image intensity can be recalculated into magnetic field
    values using a calibration procedure. At uncrossed microscope polarizers an increased $B_{z}$
    is revealed either as brighter or darker color of the image depending on the sign of $B_{z}$.
    The field resolution of the technique is of the order of a few tens of milliGauss.

Fig.\ 1 presents a set of images in the middle of a long BSCCO bar
(3300$\times$850$\times$30 $\mu m^{3}$) observed
 in the increasing in-plane field at 85K.  The in-plane field direction is indicated by the arrow.
 The images show dark lines that appear after the application of a small
 normal field $H_{z}$ = 2 Oe
 in the presence of the in-plane field. They represent enhanced values of the local magnetic induction
 and are the manifestation of Josephson vortex stacks decorated by pancake vortices. In small in-plane
 fields the decorated Josephson vortices are neither regularly spaced nor parallel to the applied field
 (Fig. 1 a, b).  In Fig.1a there is also evidence for bifurcations of the pattern as discussed in more
 detail below. Such poor ordering in small in-plane fields is caused by the weak interaction between
 the distant Josephson vortex stacks. In some regions of the sample only segments of lines were revealed
 similar to the observations in \cite {[18]} where 10 times different density of PVs along the same JVs was
 referred to transitions in PV states on a JV depending on $H_{z}$.  Buzdin
 and Baladie \cite {[20]} showed that
 PV stacks on a Josephson vortex have a long-range electromagnetic attraction due to their bending
 by JV currents. This attraction in balance with a short range repulsion provides an equilibrium distance
 between PV stacks $a_{eq}\approx 2\lambda_{ab}{\it ln} (1/\epsilon)$,
 with $\epsilon \approx(\lambda_{ab}/\lambda_{J})(B_{x}/H_{o})^{1/4}$ -a parameter characterizing
 the PV stack
 bending. Here,  $\lambda_{ab}$ is the in-plane penetration depth,
 $\lambda_{J}={\bf \gamma}s$ is the JV core size, $H_{o}=\Phi_{o}/{\bf
 \gamma}s^{2}$, $\Phi_{o}$ -the
 flux quantum, and s is the cuprate layer spacing. Above relation is valid if $a_{eq}< \lambda_{J}$.
  Note, that the
 equilibrium separation between PVs on a JV stack is controlled only by the in-plane field $B_{x}$ and
 this dependence is extremely weak. In small $H_{z}$ the available amount of pancakes is not sufficient
 to fill all JV stacks with the optimal PV density. In this case pancakes can gather (due to the long
 range character of attraction) into dense clusters where PVs are separated
 by $a_{eq}$ leaving the other
  parts of JV empty. This could explain our observations of segments of lines in the decoration patterns
  if to admit that the access of PVs is restricted in the areas where fragmented JV lines are revealed.
  In support of such situation we will show below that the inhomogeneous penetration of the normal
  field is a common feature revealed in all studied BSCCO crystals.

With increasing $H_{x}$ the pattern regularity improves. The PV
lines orient along the field, become more periodic, and the period
decreases as $H_{x}$ increases (Fig.\ 1c). At larger $H_{x}$  the
MO contrast drops but the regularity of the JV stacks is well
revealed at differences of images taken at $H_{z}$ =0 and 2 Oe in
the same $H_{x}$  (Fig.\ 1d). Such a procedure reduces the optical
noise and allows to observe much weaker features.

The improvement of the PV line pattern with $H_{x}$  can be explained by increasing the number
of PV/JV intersections and the equilibrium density of PV stacks ($\sim
1/a_{eq}$). As it was calculated
in \cite {[9]} the intersection of a Josephson vortex and a pancake line produces a negative (attractive)
contribution to the energy of both vortices due to the shift of PVs by the Lorentz force of JV
 currents. The main contribution to the crossing energy $E_{\times}$  comes from pancakes nearest to the
 JV and is given by
 \begin{eqnarray}
 E_{\times}= -2.1(\Phi_{o}^{2}/4\pi^{2}{\bf
 \gamma}^{2}s)\ln(3.5\lambda_{J}/\lambda_{ab})
 \nonumber
\end{eqnarray}
 Clearly the total
 gain of energy $\Delta E_{cr}$
  will increase with the number of intersections $N \sim 1/d$ where
  $d=[2\Phi_{o}/ {\bf \gamma}H_{x}]^{1/2}$ is the distance
  between JVs in a stack along the {\bf c}-axis. Thus the energy gain of the PV line crossing the JV
  stack compared to the energy of the noncrossing PV line $E_{pv}=
  (\Phi_{o}/4\pi \lambda_{ab})^{2}\ln (\lambda_{ab}/\xi_{ab})$ will
  be
  \begin{eqnarray}
  \Delta E_{cr}/ E_{pv}= -\frac{8.4(\lambda_{ab}/\lambda_{J})^{2}}{\ln
  (\lambda_{ab}/\xi_{ab}) \ln (3.5\lambda_{J}/\lambda_{ab})}(s/d)
   \nonumber
\end{eqnarray}
 (see
  also \cite {[11]}). Taking for
   BSCCO $\lambda_{ab}(T=0)=2000\AA $,  $\xi_{ab}(0)=30\AA $,
   $s=15\AA $, and ${\bf \gamma}=500$, one gets  $|\Delta E_{cr}|/ E_{pv}=0.055(s/d)$
   at low T.

The energy of the electromagnetic coupling between PV lines at the equilibrium
distance $a_{eq}$ on a JV stack \cite {[20]} normalized to the energy of noninteracting lines gives
 \begin{eqnarray}
\Delta E\approx \frac{-2(s/d)[1/\ln (\lambda_{ab}/\xi_{ab})] }
{\left[(\lambda_{J}/\lambda_{ab})\ln (\lambda_{J}/\lambda_{ab})
\ln[\sqrt{s/d}(\lambda_{J}/2\lambda_{ab})\ln
(\lambda_{J}/\lambda_{ab})]\right]^{2}} \nonumber
\end{eqnarray}
For comparison with $\Delta E_{cr}$(at low T) we write this ratio in a similar form
 $|\Delta E_{em}| / E_{pv} \sim 0.0017(s/d)$ at  $H_{x}$=1 Oe and
 $\sim 0.0039(s/d)$ at $H_{x}$=100 Oe (the effect of $(s/d)$
 under the $\ln$  is not large).  This shows that the major gain of energy due to the decoration
 of JV stacks by pancakes is provided by $\Delta E_{cr}$ and $\Delta E_{em}$ gives only minor contribution.
 Formally it is clear from the ratio
  \begin{eqnarray}
  \Delta E_{em}/\Delta E_{cr}\approx  \frac{4
 (\lambda_{ab}/a_{eq})^{2}}{\ln (\lambda_{J}/
  \lambda_{ab})}
   \nonumber
\end{eqnarray}

As follows from the above formulas, the {\it relative
contribution} of interactions in the PV line energy should
strongly increase near $T_{c}$ due to the temperature dependence
of $\lambda_{ab}$. Both $E_{\times}$ and $E_{em}$  are only weakly
(logarithmically) dependent on $\lambda_{ab}$ and thus on T.
However, the reference energy of a noninteracting PV line
decreases near $T_{c}$ as $\sim \lambda_{ab}^{2}$, which strongly
increases the relative effects of $E_{\times}$ and $E_{em}$.
Actually, these effects should logarithmically diverge at
$T_{d}$=86.5K, where $\lambda_{ab}(T)=
\lambda_{ab}(0)/[1-(T/T_{c})^{2}]^{1/2} \rightarrow \lambda_{J}$
(here $T_{c}$=90K).
 However, the theory is developed for $\lambda_{ab}< \lambda_{J}$ and it becomes inapplicable near $T_{d}$. We imaged
  decorated Josephson stacks in the temperature range from 60 K to 88 K. The patterns were
  poor at lower T and at T$>$87K, and the cleanest patterns were observed around 85 K.
  This is in accordance with the increasing role of $E_{\times}$ and
  $E_{em}$ at larger temperatures and
  successive instability of the crossing lattice configuration when
  $\lambda_{ab} \geq \lambda_{J}$. Note that the MO
  contrast of the pattern should decrease with temperature due to the decrease of equilibrium
  decorating PV density $1/a_{eq}\sim 1/\lambda_{ab}$. Another factor that could facilitate in the formation of
  optimum pattern at $\sim$85K is the balance between pinning restricting the mobility of pancakes
  at lower temperatures and thermal disordering washing out the vortex lattices at higher temperature.
  We note that in HTS at high T effects of correlated pinning (on extended defects such as
  twins or columnar defects) become more effective. This would extend the limits of stability
  of the PV/JV decoration patterns. These patterns (although with a weak contrast and only at
  larger fields) have been revealed even at T=88K (see Fig.2c,d).

  At a given field $H_{x}$ the only parameter controlling the period of the stacks of Josephson vortices is
  the anisotropy {\bf$\gamma$}. This was used in \cite {[18]} for extracting {\bf$\gamma$} from the decoration patterns observed in
  relatively small area ($\sim$25$\times$25 $\mu m$).  Our observations in larger
  regions ($\sim$500$\times$500 $\mu m^{2}$) revealed
  that the period can vary across the sample even in high in-plane fields and well developed patterns
   as is shown in Fig. 2. This effect can have several reasons. First, there can be local changes of the
   anisotropy associated with compositional and structural variations in the sample. It is typically
   observed in BSCCO that irregularities of the crystal structure show up at relatively high temperatures
   due to "balloons" of increased $B_{z}$ formed at them \cite {[19],[20]}. We observed that at small normal fields
    $B_{z}$ inhomogeneities in our samples serve as seeding areas from which PVs spread along the JV
     stacks. Similar channeling of PVs along JVs was reported in
     \cite {[18]}. In the vicinity of the increased
     $B_{z}$ spots the decoration pattern showed smaller periods at all fields. This can be associated with
     compositional variations of the anisotropy (e.g. increased oxygen
     concentration should reduce {\bf$\gamma$} \cite {[23]}
      and thus the period). However, in some areas the difference in period compared to the neighboring
      regions was found to disappear at increasing field.  This could be due to a possibility of two different
      JV configurations as shown in Fig.3, which have different relations for the stack periods ($D$) and
     distances between JVs in the stack ($d$): $D_{1}=( \sqrt{3}{\bf \gamma}\Phi_{o}/2H_{x})^{1/2}$,
      $d_{1}=(2\Phi_{o}/  \sqrt{3}{\bf \gamma}H_{x})^{1/2}$, and
      $D_{2}=({\bf \gamma}\Phi_{o}/ 2\sqrt{3}H_{x})^{1/2}$,
      $d_{2}=(2\sqrt{3}\Phi_{o}/ {\bf \gamma}H_{x})^{1/2}$, respectively.
      Within the anisotropic London model such vortex configurations have
       the same energy when the field is along the $ab$ plane \cite
       {[24]}. This degeneracy, however, is lifted and the
       first configuration becomes preferential when the perpendicular
       field is applied \cite {[24]}. In the case of the
       layered systems in the crossing fields the difference in the energy of the configurations will be due
        to different number of PV/JV intersections. For one PV line intersecting a JV stack this difference
         will be $\Delta E_{cr} \sim E_{\times} \cdot (1/d_{1}-1/d_{2}$) which
         also gives preference to the first configuration
         ($d_{1}<d_{2}$).
         However, this preference is very small. Assuming the equilibrium distances between PV lines on
         JV stacks the total number of intersections per unit volume in
         each configuration will be $N_{cr} \sim 1/d \cdot 1/D \cdot 1/a_{eq}$.
         In this case the difference in the gain due to the crossing energy between
         them totally disappears ($d_{1}D_{1}= d_{2}D_{2}$).
         The average value of $B_{z}$ will be slightly different for the two configurations. However, due to the
          inhomogeneities in the samples tuning of the local $\langle
          B_{z} \rangle $ can be easily realized. Therefore one can expect
           that these configurations could coexist in the samples and transform from one to another causing
            changes of the decoration pattern period. In the increasing in-plane field both distances between
             the stacks, D, and spacing between JVs along the $c$-axis, $d$, should decrease. For tuning $D$ Josephson
              vortices easily slide along the cuprate planes. However, to tune $d$ they have to traverse the planes which
              requires overcoming a considerable energetic barrier and makes this process much slower. It can be
              probably done only by formation of double kinks or kinks coming from the edges and their recession
               along the planes as in the case of dislocations moving from
               one crystal plane to another \cite {[25]}. It is
                reasonable to suggest that the system is not in the equilibrium state at all fields but transforms through
                a set of metastable states admitting a mixture of the above two possible JV configurations and transitions
                between them. A whole series of other JV configurations (with different D/d ratios) transforming from
                one to another in small $H_{x}$ were predicted for the situation when JV vortices can not move along the
                $c$ axis at all \cite {[26]}. Stepwise transitions between different JV configurations in higher fields were
                discussed recently in \cite {[27]}.  A possible confirmation
                of the above scenario is the $D^{-2}(H_{x})$ dependence
                shown in Fig.\ 4. Minimum, maximum and average periods of the decoration pattern plotted on this
                graph were measured over a large area (500$\times$500 $\mu m^{2}$) in the middle of the crystal. The data
                definitely show a number of steps which can be associated with impeding of the JV system tuning and
                possibly transitions between different JV lattice configurations. At the same time, the overall changes
                of the decoration pattern period is well described by a
                linear dependence $D^{-2}$ vs. $H_{x}$  and give
                reasonable values of the anisotropy from 570 to 700.

    Another kind of variation of the pattern period was observed at moderate in-plane fields and at higher
    temperatures when areas showing the same anisotropy in larger $H_{x}$   unexpectedly reveal periods differing
    by a factor of two.  Fig. 5a, b show such an area where the distance between decorated JV stacks increases
    with increasing field in contradiction to theory.  However, the pattern in nearby areas (Fig. 5c) where the
    period decreases with increasing $H_{x}$   in a regular way displays half of the period seen in the anomalous areas.
    This indicates that the large period reveals only each second JV stack.  The possibility of the formation of PV
     chains at integer numbers of JV periods was considered recently by
     Buzdin and Baladie \cite{[20]}.  They show that
      the equilibrium density of PV lines decorating JV stack
      ($1/a_{eq}$) is practically field independent (it is a very
      weak function of $H_{x}$   only). So at low $B_{z}$   there are not enough pancakes to decorate all the JV stacks with
      this density and a superstructure of decorated and non-decorated
      stacks develops. With increasing $B_{z}$ the
      density of chains increases and finally all JV stacks become occupied with PV.  In large $B_{z}$ pancakes start
      spreading into the volume between the JVs. Obviously, in our samples inhomogeneities of the PV concentration
      caused by crystal imperfections determine the domains where local conditions for superstructures are provided.
      We already exploited the argument of equilibrium PV spacing to explain the observation of segmented
      decoration lines. To explain the commensurate periodic patterns one has to account also for the long range
      repulsion between pancakes on different stacks. This could be due
      to the Pearl's interaction \cite {[28]} through the
      stray fields of pancakes at the sample surface. Although these interactions are long range but they are weak
      and their effect should be noticeable only in favorable conditions. Probably this is the reason why the
      superstructure was observed only at high temperatures and only in some parts of the sample.

  The picture of the field transformations of the JV lattice discussed above suggests that there should be defects
  in the JV stack arrangement. In particular, we observed interconnects and ``dislocations'' in the regular decoration
  line arrays as illustrated in Fig.6. It is known that JV lines do not split or terminate inside a superconductor.
  Therefore, the decoration pattern reveals the migration of Josephson vortices between different stacks.
  A corresponding scheme of JV kinks visualized due to PV stacks of different length is shown at the bottom of Fig.6.
  The kinks can form at different depth from the surface and still they will be decorated by PV stacks if this depth
   is not too large. On the left scheme two possible variants of JV travelling between the stack is shown but only one
   of them corresponds to the pattern in Fig.6a. Note, that in either case the kinks in neighboring stacks are located
   between different cuprate planes. This means that defects in the decoration pattern reveal also defects in ordering
   of Josephson vortices in the stacks along the {\bf c}-axis. Comparison of Fig.6b and c shows that JV kinks shift along
   the direction of $H_{x}$ at changing field. As a result the defects are removed from the decoration pattern at large in-plane fields.

Difference images presented in Figs.6b and c illustrate one more important feature of the JV behavior.
The first picture corresponds to the increase of $H_{x}$  by 2 Oe and the second one to the increase of $H_{z}$  by the
same 2 Oe. Old positions of JV stacks are revealed as brighter lines and new positions are seen as dark lines
(the direction of motion is shown by arrow on the top). It is easy to see
that at T$\sim$85K and $H_{x}\sim$ 25 Oe the
 change of $H_{z}$  shifts JVs by nearly the same distance as a similar change of $H_{x}$. This is a direct evidence
 of the strong coupling between the two systems of vortices. It also reveals a possibility to move Josephson
 vortices either directly acting on them by the in-plane field or by shifting coupled to them pancake vortices
 by the normal field.

Another illustration of strong interactions between the two components of magnetic flux in BSCCO
is the effect of $B_{z}$  inhomogeneities on bending JV stacks. Fig.7a-b presents two successive images taken
 at increasing $H_{z}$  from 5 to 8 Oe in a sample field cooled to 83K
 in $H_{x}$ =18.5 Oe. Dark spots on the
 pictures correspond to increased $B_{z}$  at structural imperfections of the crystal as discussed before.
 Dark lines are JV stacks decorated by PVs. Upon ramping $H_{z}$  up the increased $B_{z}$  region in the
  left bottom corner expands and shifts decorated JV stacks considerably bending them. Fig.7c shows that
   Josephson vortices can also strongly bend during their evolution in the in-plane field when they cross
   the $B_{z}$  inhomogeneities. This picture presents the difference of images at $H_{x}$ =12 and 14 Oe. A rounded
   area in the middle of the picture is the spot of increased $B_{z}$. It is not dark because at original images $B_{z}$
   in this spot did not change. Here JV lines are not revealed because they did not move at changing $H_{x}$ being
   pinned by the large density of PVs. The break of angle of the Josephson vortices in this spot is well seen.

In general, the JV stacks in the crossing fields are clearly seen on the background of the inhomogeneous
$B_{z}$  distribution as shown on a panoramic MO picture near the corner of a BSCCO crystal in Fig.8. They
form an intricate pattern of curved lines (it straightens up at larger $H_{x}$ ) which can be a sensitive tool for
studying crystal imperfections and spatial anisotropy variations.

\section{Conclusions}

We studied magneto-optically the behavior of the decoration of Josepshon vortices by pancake vortices in the single
 crystals of BSCCO. The decoration pattern turns out to be robust in a wide range of temperatures and fields.
 Simultaneous imaging of a large area accessible by the MO technique allows to observe interactions of the Josephson
 vortices with sample imperfections and through the measurements of JV stack periods to map spatial variations of
 the anisotropy of the sample. The experiment reveals a strong coupling between pancake and Josephson vortices
 which shows up through the motion of both flux components under the direct action on one of them and pinning
 and bending of JV lines at the inhomogeneities of the pancake density ($B_{z}$ ). At larger temperatures the quality of
 the decoration pattern improves which can be referred to (a) the relative increase of the energy of PV/JV interactions
 compared to the energy of noninteracting pancakes and (b) the increase of the role of correlated pinning of pancakes
  on the nearly 2D JV stacks. Possible evidences of the coexistence of two different JV lattice configurations are observed.
  Also intrinsic defects of the JV lattice like ``interconnects'' and ``dislocations'' are revealed that are associated with wandering
  of the Josephson vortices between neighboring stacks. At small fields commensurate patterns with each second JV
  stack decorated by pancakes are found in accordance with recent theoretical predictions.

 \section*{Acknowledgments}
The authors thank A.\ I.\ Buzdin and S.\ Bending for stimulating
discussions and presentation of his results prior to publication.
The work was supported by the U.S.DOE, BES-Materials Sciences,
under Contract \#W-31-109-ENG-38.

\newpage
  \begin{figure}
\epsfxsize=3.3in
 \epsffile{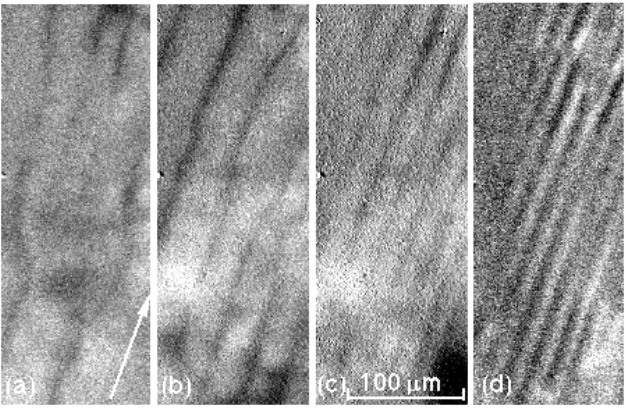}
 \vspace *{0.1cm}
 \caption{Lines of increased $B_{z}$ (dark) at stacks of Josephson vortices at changing in-plane field.
  $H_{z}$=2 Oe, $H_{x}$ for (a) to (d): 3, 12, 24, and 42 Oe. A ratio of
  images ($\uparrow$2,42)/(0,42) in (d) reveals better
   the positions of JV stacks. T=85K.}
\label{Fig-1}
\end{figure}

\vspace*{0.5cm}
   \begin{figure}
 \epsffile{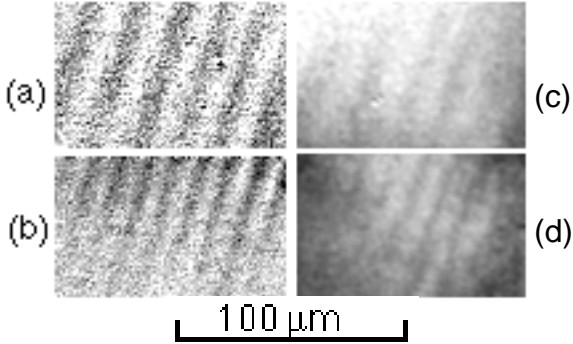}
 \vspace*{0.1cm}
 \caption{Different periods of MO patterns in different spots distant by ~600 $\mu$m ((a) and (b) at 84K,
  (c) and (d) at 88K) in a BSCCO crystal. Ratios of images ($H_{z}$,
  $H_{x}$)= ($\uparrow$2, 40)/(0,40) Oe are shown.}
 \label{Fig-2}
 \end{figure}

    \begin{figure}
    \leftline{
 \epsfxsize=3.3in
 \epsffile{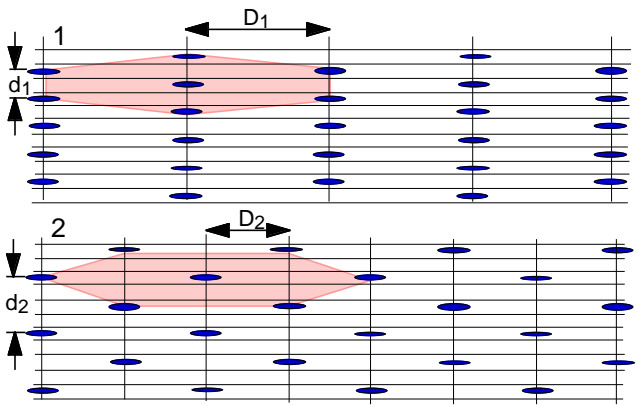}
 }
 \vspace*{0.3cm}
   \caption{Schematic for two different Josephson vortex configurations in a layered
   superconductor (for ${\bf \gamma} \sim$ 5) which are degenerate in energy
   at H $\parallel$ cuprate planes.
   For BSCCO the picture should be stretched along X and compressed along
   Y $\sim $ 10 times.}
 \label{Fig-3}
 \end{figure}

\begin{figure}
    \epsfxsize=3in
 \epsffile{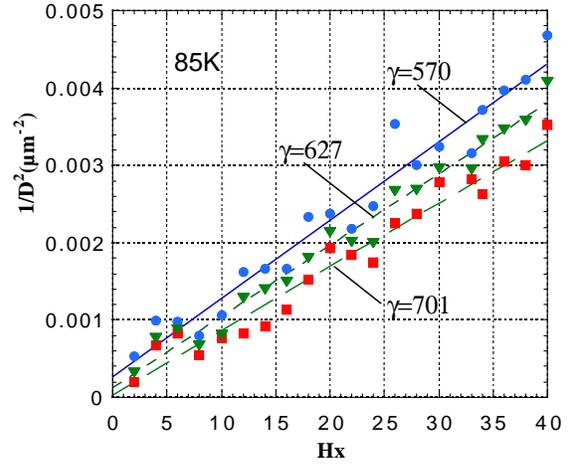}
 \vspace*{-2.5cm}
   \caption{Field dependence of the decoration pattern period. Linear fit for $1/D^{2}-H_{x}$ gives
   estimates for the anisotropy for min, max, and average periods as shown on the graph.}
 \label{Fig-4}
 \end{figure}

 \begin{figure}
 \epsffile{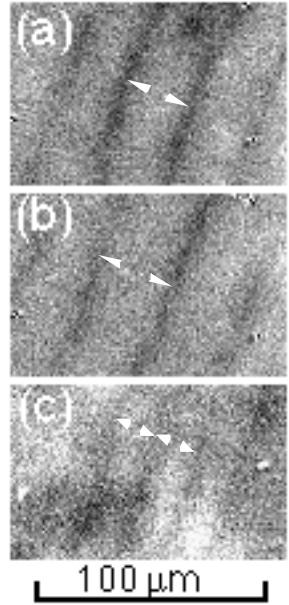}
 \vspace*{0.3cm}
   \caption{Increase of the decoration pattern period (marked by arrows) at increasing
   in-plane field from 15 Oe in (a) to 21 Oe in (b). (c)  Twice smaller period revealed in
    the neighboring area at 21 Oe. T=83K, $H_{z}$=2 Oe.}
 \label{Fig-5}
 \end{figure}

  \begin{figure}
 \epsfxsize=3in
 \epsffile{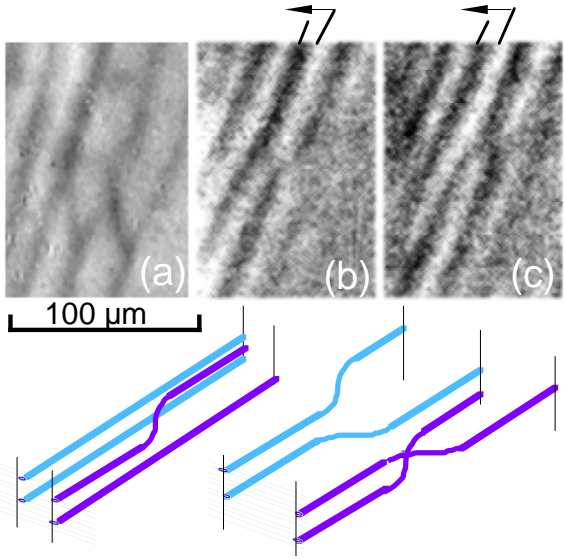}
 \vspace*{0.3cm}
   \caption{Interconnects and dislocation like structures in the JV decoration patterns.
    (a) Difference of MO images at ($H_{z},H_{x}$)=(2, 18.5) and (0,18.5) Oe, T=83K.
    (b) Difference of images at ($H_{z},H_{x}$))=(0, 26) and (0, 24) Oe,  JVs carry PVs
    captured after previous application of $B_{z}$=2 Oe, T=85K. (c) Difference of images
    at ($H_{z},H_{x}$))=(2, 26) and (0,26) Oe, T=85K. Schematics of kinks at Josephson vortices
    rerouting them between neighboring stacks are shown in the bottom. Left scheme
    corresponds to (a) and right to (b-c). Vertical lines on the schematics mark positions
    of JV stacks. Horizontal lines show cuprate planes.}
 \label{Fig-6}
 \end{figure}

 \begin{figure}
 \epsfxsize=6in
 \epsffile{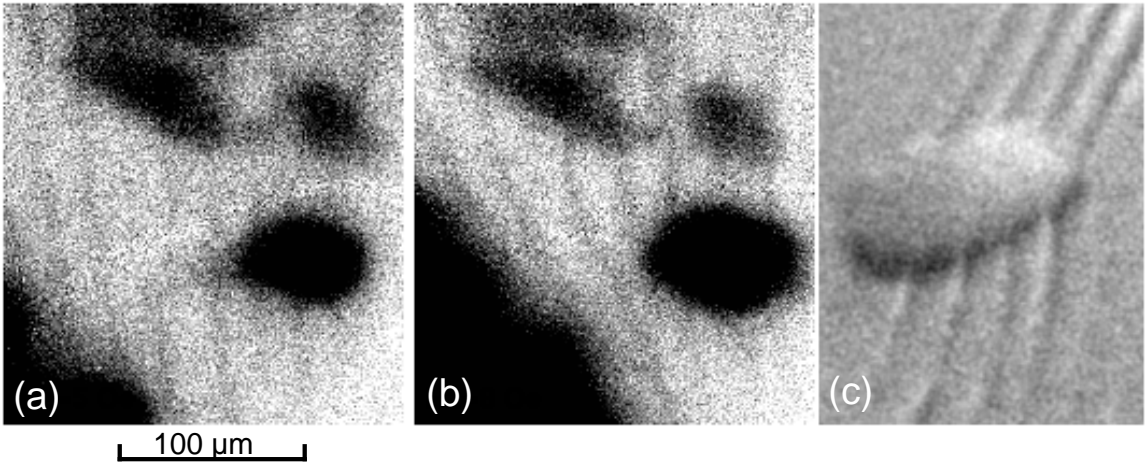}
 \vspace*{0.5cm}
   \caption{(a-b)Bending of Josephson vortices (dark lines ~ in y-direction)
   at spots of increased $B_{z}$ (dark). (a) $H_{z}$=5 Oe, (b) $H_{z}$=8, T=83K. The sample is
   cooled in $H_{x}$=18.5 Oe. (c) The difference of MO images taken at
   changing $H_{x}$ from 12 to 14 Oe near the spot of increased $B_{z}$ (rounded area in the center). $H_{z}$=10 Oe,
   T=80K. JVs have shifted from bright to dark positions.}
 \label{Fig-7}
 \end{figure}

 \begin{figure}
 \epsfxsize=3in
 \epsffile{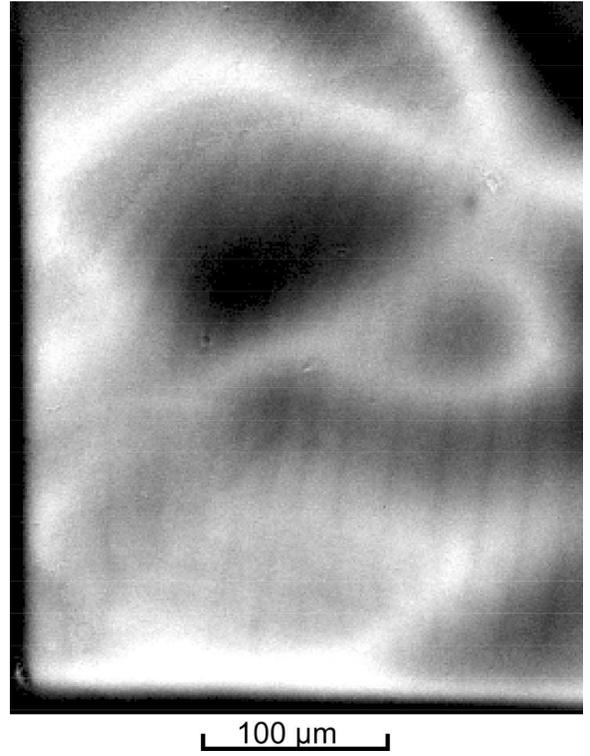}
 \vspace*{0.3cm}
   \caption{Josephson vortex stacks (dark lines in vertical direction) visualized
    at $H_{z}$=22 Oe $H_{x}$ =20 Oe , T=83K. Brightness of the image corresponds to the
    local $B_{z}$ in the sample. Left and bottom sides of the picture are crystal edges.}
 \label{Fig-8}
 \end{figure}


\begin {references}
\bibitem{[1]} A.\ Buzdin and D.\ Feinberg, J.\ Phys.\ {\bf 51}, 1971 (1990).
\bibitem{[2]} J.\ R.\ Clem, Phys.\ Rev.\ B {\bf 43}, 7837 (1991).
\bibitem{[3]} S.\ N.\ Artemenko and A.\ N.\ Kruglov, Phys.\ Lett.\  {\bf 83} 485 (1990).
\bibitem{[4]} P.\ H.\ Kes, J.\ Aarts, V.\ M.\ Vinokur, and C.\ J.\  van der Beek ,
Phys.\ Rev.\ Lett.\  {\bf 64} , 1063 (1990).
\bibitem{[5]} S.\ Theodorakis, Phys.\ Rev.\ B {\bf 42}, 10172 (1990).
\bibitem{[6]} L.\ N.\ Bulaevskii, M.\ Ledvij, and V.\ G.\ Kogan, Phys.\ Rev.\ B {\bf 48}, 366 (1992).
\bibitem{[7]}  M.\ Bencrouda and M.\ Ledvij, Phys.\ Rev.\ B {\bf 51}, 6123 (1995).
\bibitem{[8]}  L.\ N.\ Bulaevskii, M.\  Maley, H.\ Safar, and D.\ Dominguez, Phys.\ Rev.\ B  {\bf 53}, 6634 (1996).
\bibitem{[9]} A.\ E.\ Koshelev, Phys.\ Rev.\ Lett.\  {\bf 83}, 187 (1999).
\bibitem{[10]} S.\ E.\ Savel'ev, J.\ Mirkovich, and K.\ Kadowaki, Phys.\ Rev.\ B {\bf 64}, 094521 (2001).
\bibitem{[11]}M.\ J.\ W.\ Dodgson, cond-mat/0201197 (2002).
\bibitem{[12]}C.\ A.\ Bolle, P.\ L.\ Gammel, D.\ G.\ Grier, C.\ A.\ Murray, and D.\ J.\ Bishop,
Phys.\ Rev.\ Lett.\  {\bf 66}, 112 (1991).
\bibitem{[13]} I.\ V.\ Grigorieva, J.\ W.\ Steeds, G.\ Balakrishnan, and D.\ M.\ Paul,
Phys.\ Rev.\ B {\bf 51}, 3765 (1995).
\bibitem{[14]} A.\ I.\ Buzdin and A.\ Yu.\ Simonov, JETP Lett.\  {\bf 51}, 191 (1990).
\bibitem{[15]} A.\ M.\ Grishin, A.\ Yu.\ Martynovich, and S.\ V.\ Yampolskii, Sov.\ Phys.
JETP {\bf 70}, 1089 (1990).
\bibitem{[16]} V.\ G.\ Kogan, N.\ Nakagawa, and S.\ L.\ Thiemann, Phys.\ Rev.\ B {\bf 42}, 2631 (1990).
\bibitem{[17]} D.\ Huse, Phys.\ Rev.\ B {\bf 46}, 8621 (1992).
\bibitem{[18]}  A.\ Grigorenko, S.\ Bending, T.\ Tamegai, S.\ Ooi, and M.\ Henini, Nature
{\bf 414}, 728 (2001).
\bibitem {MatsudaSci02}T.\ Matsuda, O.\ Kamimura, H.\ Kasai, K.\ Harada, T.\ Yoshida,
T.\ Akashi, A.\ Tonomura, Y.\ Nakayama, J.\ Shimoyama, K.\ Kishio,
T.\ Hanaguri, and K.\ Kitazawa, Science, \textbf{294}, 2136(2001)
\bibitem{[19]} V.\ K.\ Vlasko-Vlasov, U.\ Welp, G.\ W.\ Crabtree, and V.\ I.\ Nikitenko in {\it Physics and
  Materials Science of vortex states, flux pinning and dynamics} (eds.\ R.\ Kossowsky et al.,
  NATO Science Series E) {\bf 356}, 205-237 (Kluwer Academic Publishers, Dordrecht, Boston, London, 1999).
\bibitem{[20]}  A.\ Buzdin and I.\ Baladie, cond-mat/0110339 (2001).
\bibitem{[21]} A.\ Soibel, E.\ Zeldov, M.\ Rapoport, Yu.\ Myasoedov, T.\ Tamegai, Sh.\ Ooi,
M.\ Konczykowski, and V.\ B.\ Geshkenbein, Nature {\bf 406}, 282
(2000).
\bibitem{[22]} G.\ Yang, J.\ S.\ Abell, and C.\ E.\ Gough, Physica C 341-348, 1091 (2000).
\bibitem{[23]}  Y.\ Kotaka, T.\ Kimura, H.\ Ikuta, J.\ Shimoyama, K.\ Kitazawa, K.\ Yamafuji,
K.\ Kishio, and D.\ Pooke, Physica C {\bf 235}-{\bf 240}, 1529
(1994).
\bibitem{[24]} L.\ J.\ Campbel, M.\ M.\ Doria, and V.\ G.\ Kogan, Phys.\ Rev.\ B {\bf 38}, 2439 (1988).
\bibitem{[25]}  J.\ P.\ Hirth and J.\ Lothe, Theory of Dislocations, Wiley, New York (1982).
\bibitem{[26]} L.\ S.\ Levitov, Phys.\ Rev.\ Lett.\   {\bf 66}, 224 (1991).
\bibitem{[27]}  J.\ Mirkovic, S.\ E.\ Savel'ev, E.\ Sugahara, and K.\ Kadowaki,
Phys.\ Rev.\ Lett.\  {\bf 86}, 886 (2001).
\bibitem{[28]}  J.\ Pearl, Appl.\ Phys.\ Lett.\  {\bf 5}, 65 (1964).

\end{references}
\end {document}